\documentclass[acus]{JAC2000}

\usepackage{graphicx}

\begin{document}
\title{\flushright{PSN}\\[15pt] \centering CMD-2 DETECTOR UPGRADE}

\author{D. Grigoriev, BINP, Novosibirsk, 630090, Russia \\
on behalf of CMD-2M collaboration\thanks{CMD-2M
collaboration is listed in appendix~A. } }

\maketitle

\begin{abstract}
The project of upgrading the detector CMD-2 
is presented. The upgraded detector
is called \mbox{CMD-2M} and is going to take data 
with new collider VEPP-2000 at BINP. 
The general structure of the detector CMD-2 will
remain as is but major parameters of the detector,
such as momentum and angular resolution for 
charged particle and energy and spatial resolution
for photons, will be substantially improved. 
\end{abstract}

\section{INTRODUCTION}

\begin{figure}[htb]
\begin{center}
\includegraphics[width=0.45\textwidth]{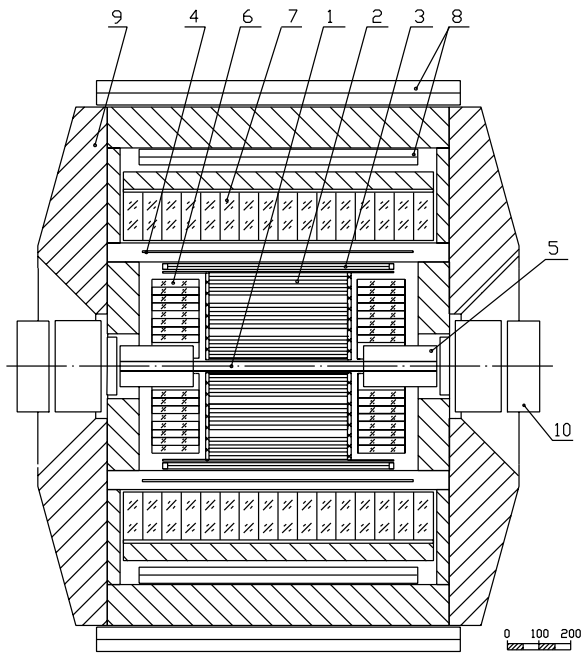}
\end{center}
  \caption{Layout of the CMD-2 detector.
1~--~vacuum chamber, 2~--~drift chamber, 3~--~Z-chamber,
4~--~main solenoid, 5~--~compensating solenoid,
6~--~endcap calorimeter, 7~--~barrel calorimeter,
8~--~range system, 9~--~flux return yoke,
10~--~storage ring lens }
\label{fig:cmd2}
\end{figure}

The general-purpose Cryogenic Magnetic Detector 
(\mbox{CMD-2})~(fig.~\ref{fig:cmd2})~\cite{cmd2}
has been running at the VEPP-2M electron-positron collider in
Novosibirsk from 1992 to 2000 studying the centre-of-mass 
energy range from 0.36 to 1.4~GeV. The total integrated 
luminosity collected is about 25~pb$^{-1}$. It allows
the study, with high precision, of many channels of 
$e^{+}e^{-}$~annihilation to hadrons and rare decays
of the light vector mesons~\cite{cmd2_res}. 

At present time a modernization of VEPP-2M 
accelerator-collider complex is in progress. 
The main part of the modernization is the creation
of the new collider VEPP-2000~\cite{vepp2000}
with centre-of-mass energy up to 2~GeV and
luminosity up to 10$^{32}$~cm$^{-2}$s$^{-1}$.

The change of the collider interaction point
design and the increase of its energy and luminosity
require the upgrade of the detector. 
The general structure of the detector will be
the same. The most expensive components of the CMD-2
detector, such as BGO and CsI crystals, as well
as substantial part of digitizing electronics
and software will be used in the CMD-2M detector.
But all systems of the detector will be upgraded
or replaced by new ones.

\section{GENERAL DESCRIPTION OF THE CMD-2M DETECTOR }

\begin{figure*}[htb]
\begin{center}
\includegraphics*[height=0.45\textwidth]{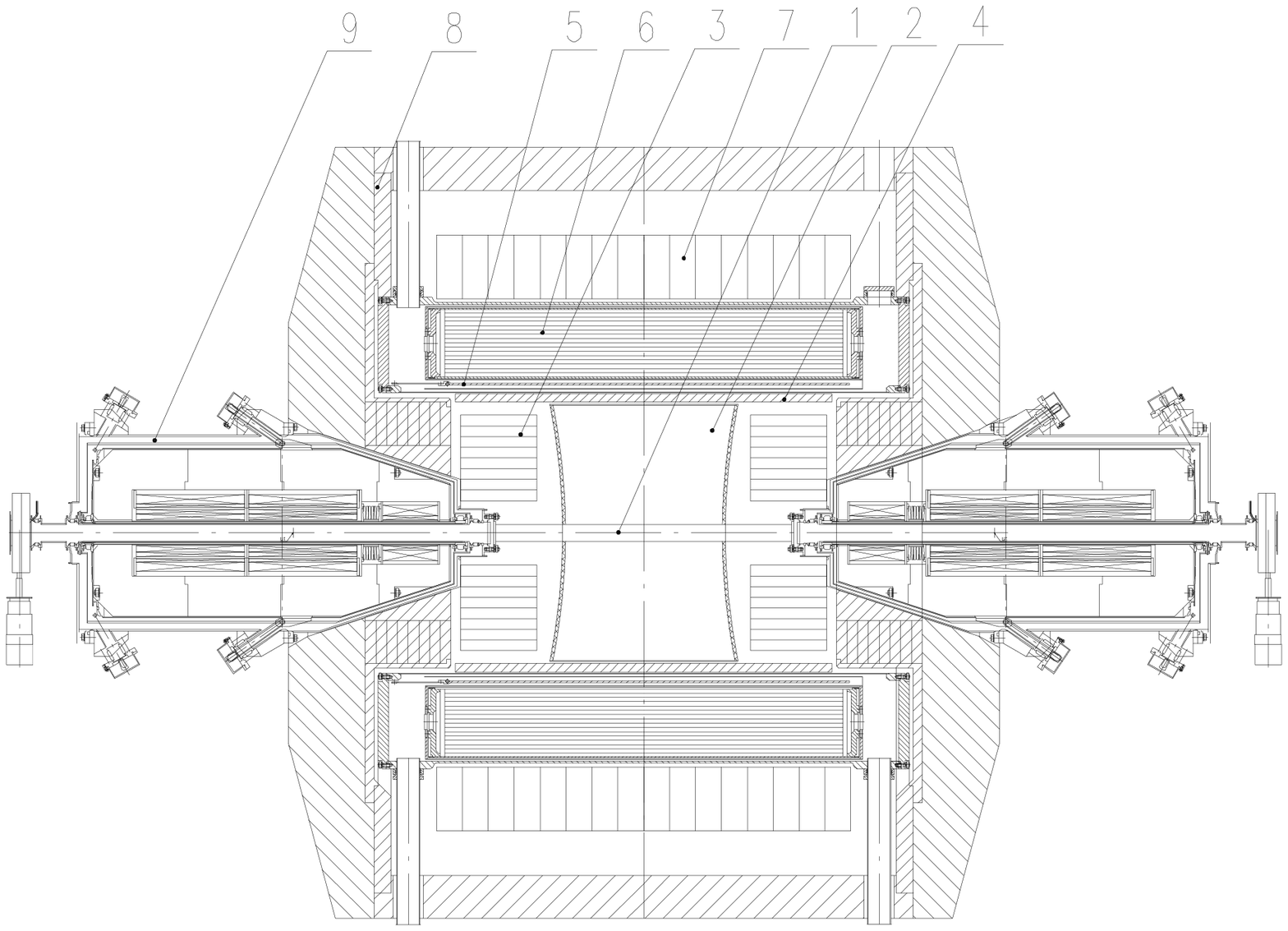}
\hfill
\includegraphics*[height=0.45\textwidth]{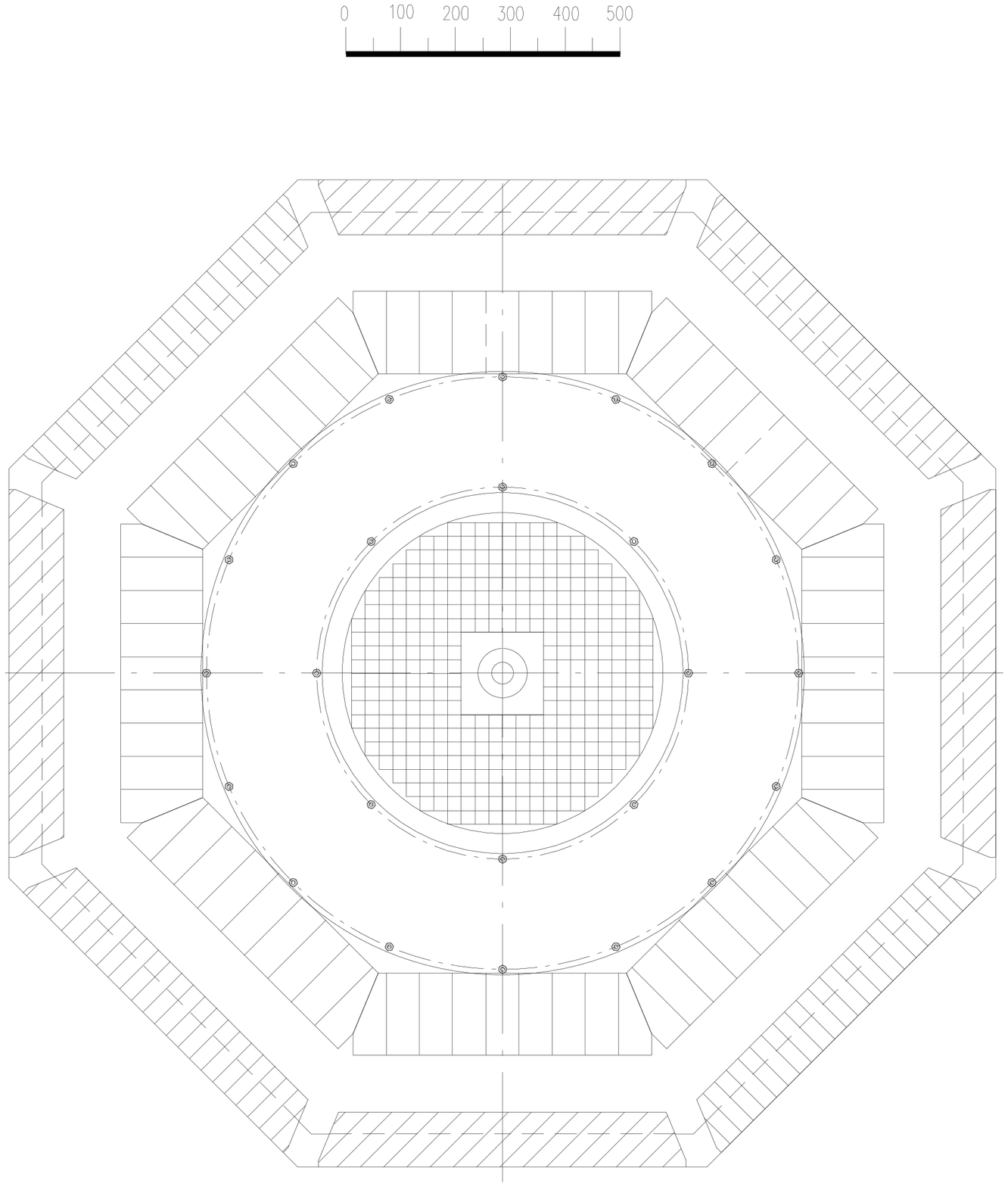}
\\
\end{center}
  \caption{ Layout of the CMD-2M detector. 
1~--~vacuum chamber, 
2~--~drift chamber, 
3~--~BGO endcap calorimeter, 
4~--~\mbox{Z-chamber},
5~--~main solenoid, 
6~--~liquid Xe barrel calorimeter,
7~--~CsI barrel calorimeter, 
8~--~flux return yoke,
9~--~collider focusing solenoids. }
\label{fig:cmd2m}
\end{figure*}

The layout of the CMD-2M detector is shown 
on fig.~\ref{fig:cmd2m}. 

The electron and positron beams collide 
at the center of the vacuum chamber~(1) 
which has an internal diameter of 34~mm. 
The central part of the vacuum chamber is 
a Be tube with
thickness of 0.77~mm~(2.1$\times$10$^{-3}$X$_{0}$).
A set of masks protects thin part of the vacuum chamber 
from synchrotron radiation.

The coordinate, angle and momentum of charged particles 
are measured by the cylindrical drift chamber~(2) with
hexagonal cell. The chamber diameter is 600~mm, its
length 440~mm. The charge division along the wires
is used for measuring the Z-coordinate of tracks.

The Z-chamber~(4) produces fast pretrigger signal for
charge trigger and measures the Z-coordinate of tracks
with high resolution.

The drift and Z-chambers are placed inside the 
superconductive solenoid~(5) with magnetic field of 1.5~T.
The total thickness of the solenoid is 0.18~X$_{0}$. 
The collider focusing solenoids~(9) with magnetic field 
of 13~T are almost
inside the detector but their influence on the magnetic
field in the volume of drift chamber is small. 

The coordinates and energy of photons
are measured by the barrel calorimeter based on liquid Xe~(6)
and CsI crystals~(7) and endcap calorimeter based on 
BGO crystals~(3). 
Both the barrel and endcap calorimeters cover a solid angle
of 0.94$\times 4 \pi$~steradians. The barrel calorimeter
is located outside the magnet. To minimize the amount
of passive matter in front of the calorimeter the liquid
Xe calorimeter and the magnet are placed into
a  common cryostat.

The iron flux return yoke~(8) is octagonal in shape with 
a length of 1660~mm and a height of 1800~mm. 
The range system, based on plastic scintillator,
is located outside the yoke.

The main parameters of the CMD-2M detector are listed 
in table~\ref{tab:cmd2m} for comparison with the CMD-2 detector.

\begin{table*}[tbh]
\begin{center}
\caption{Main parameters of CMD-2 and CMD-2M detectors.}
\label{tab:cmd2m}
\vspace{0.5 cm}
\begin{tabular}{|p{0.15\textwidth}|p{0.35\textwidth}|p{0.35\textwidth}|}
\hline
{\centering \vspace{1mm} \bf System \\} & 
{\centering \vspace{1mm} \bf CMD-2 \\} & 
{\centering \vspace{1mm} \bf CMD-2M \\} \\
\hline
{\centering \vspace{1mm} Drift \\ chamber \\} & 
{\centering \vspace{1mm} 512  sensitive wires \\ 
 $\sigma_{R-\phi}=250~\mu$m, $\sigma_{Z}=5$~mm, \\
 $\sigma_{\theta}$=15$\cdot 10^{-3}$, $\sigma_{\phi}=7\cdot 10^{-3}$, \\
 $\sigma_{dE/dx}$=0.2$\cdot E$ \\[-15mm]}  &
{\centering \vspace{1mm} 1218 sensitive wires \\
 $\sigma_{R-\phi}=140~\mu$m, $\sigma_{Z}=2$~mm, \\
 $\sigma_{\theta}$=7$\cdot 10^{-3}$, $\sigma_{\phi}=4\cdot 10^{-3}$, \\
 $\sigma_{dE/dx}$=0.15$\cdot E$ \\} \\
\hline
{\centering \vspace{1mm} Z-chamber \\}& 
\multicolumn{2}{|p{0.7\textwidth}|}{\centering \vspace{1mm} Double 
layers proportional chamber
 with cathode strips\\
anode wires are combined to 2$\times$32  sectors,
number of cathode strips - 512 \\
$\sigma_{Z}=250 \div 1000~\mu$m , $\sigma_{t}$=5~ns \\ } \\
\hline
{\centering  \vspace{1mm} Barrel \\ Calorimeter \\}& 
{\centering \vspace{1mm} 892 CsI crystals in 8 octants \\
readout PMT \\
\mbox{}\\
thickness 8.1 $X_{0}$ \\
$\sigma_{E}/E=8$\%, 
$\sigma_{\theta,\phi}=0.03 \div$0.02~rad \\
 at $E_{\gamma}=100\div$700~MeV\\[-0.5\baselineskip]}&  
{\centering \vspace{1mm} 1152 CsI crystals in 8 octants \\
readout Si photodiodes \\
400~l~LXe \\
thickness 5(LXE)+8.1(CsI)=13.1~$X_{0}$ \\
$\sigma_{E}/E=4.7\div3$\%, 
$\sigma_{\theta,\phi}$=0.005~rad \\
 at $E_{\gamma}=100\div$900~MeV\\[-0.5\baselineskip]} \\
\hline
{\centering \vspace{1mm} Endcap \\ Calorimeter \\}&
{\centering \vspace{1mm} 680 BGO crystals in 2 endcaps \\
readout vacuum phototriodes \\
thickness 13.4 $X_{0}$ \\
$\sigma_{E}/E=8\div4$\%, 
$\sigma_{\theta,\phi}=0.03 \div$0.02~rad \\
 at $E_{\gamma}=100\div$700~MeV\\[-0.5\baselineskip]}&
{\centering  \vspace{1mm} 680 BGO crystals in 2 endcaps \\
readout Si photodiodes \\
thickness 13.4 $X_{0}$ \\
$\sigma_{E}/E=8\div3.5$\%, 
$\sigma_{\theta,\phi}=0.03 \div$0.02~rad \\
 at $E_{\gamma}=100\div$900~MeV\\[-0.5\baselineskip]} \\
\hline
{\centering \vspace{1mm} Range system \\[-0.5\baselineskip]}&
{\centering \vspace{1mm} Streamer tubes, 2 double layers, 
$\sigma_{Z}$=5~cm \\[-0.5\baselineskip]}&
{\centering \vspace{1mm} Plastic scintillator counters, 
$\sigma_{t}<$1 ns \\[-0.5\baselineskip]} \\
\hline
{\centering \vspace{1mm} Superconductive \\ solenoid \\}&
{\centering \vspace{1mm} Magnetic field 1 T,
thickness 0.38 X$_{0}$ \\}&
{\centering \vspace{1mm} Magnetic field 1.5 T,
thickness 0.18 X$_{0}$ \\} \\
\hline
\end{tabular}
\end{center}
\end{table*}

\section{Drift Chamber}

The drift chamber of the CMD-2 detector has jet-like 
structure~\cite{dc}. At energies above 1~GeV the event 
multiplicity increases substantially. Events
with at least 4~tracks have largest cross sections.
To study those events a drift chamber with
a uniform, relatively small cell, is well suitable.
The cell has a hexagonal shape with a diagonal
distance of 17~mm. Also this structure
allows for high reconstruction efficiency for particles, 
decaying inside the sensitive volume, that is important
for K$_{S}$-meson reconstruction.

The number of cells for chosen geometry is~1218. The sensitive
wires of 15~$\mu$m diameter are made with gold-plated
\mbox{W-Re}~alloy. The field wires of 80~$\mu$m diameter are 
made with gold-plated titanium. The ratio between numbers
of sensitive and field wires is~1:2. 

The gas mixture is Ar:iC$_{4}$H$_{10}$~(80:20), 
and th gas gain
coefficient is about 10$^{5}$ at 2~kV applied voltage.
The electric field strength on the surface of 
the field wires is less than 20~kV/cm which is safe from 
the point of view of aging. Maximum drift time is 600~ns.

\begin{figure}[tb]
\begin{center}
\includegraphics[width=0.45\textwidth]{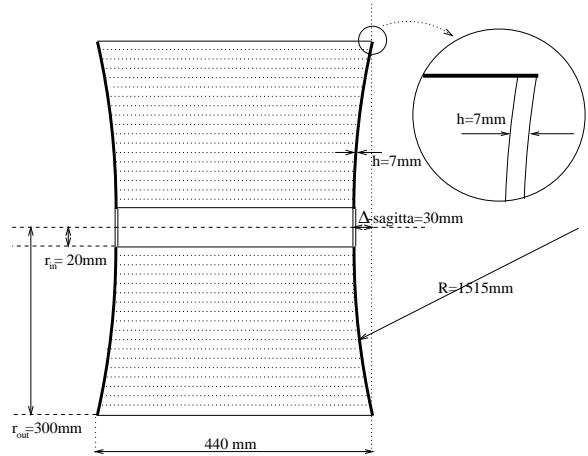}
\end{center}
  \caption{Layout of the drift chamber.}
\label{fig:dc_constr}
\end{figure}

The frame of the chamber is made from carbon fibers. 
To minimize the amount of passive matter in front of 
the endcap calorimeter the flanges are spherical in shape
with a radius of curvature of 1515~cm~(fig.~\ref{fig:dc_constr}).
The flanges thickness is 7~mm, and the outer wall thickness
2~mm. The inner wall is made from 0.2~mm kapton. 
The chamber thickness for 90~degrees tracks is 0.01~X$_{0}$.
The thickness of passive matter in front of
the endcap calorimeter is 0.04~X$_{0}$.

\begin{figure}[tbh]
\begin{center}
\includegraphics[width=0.45\textwidth]{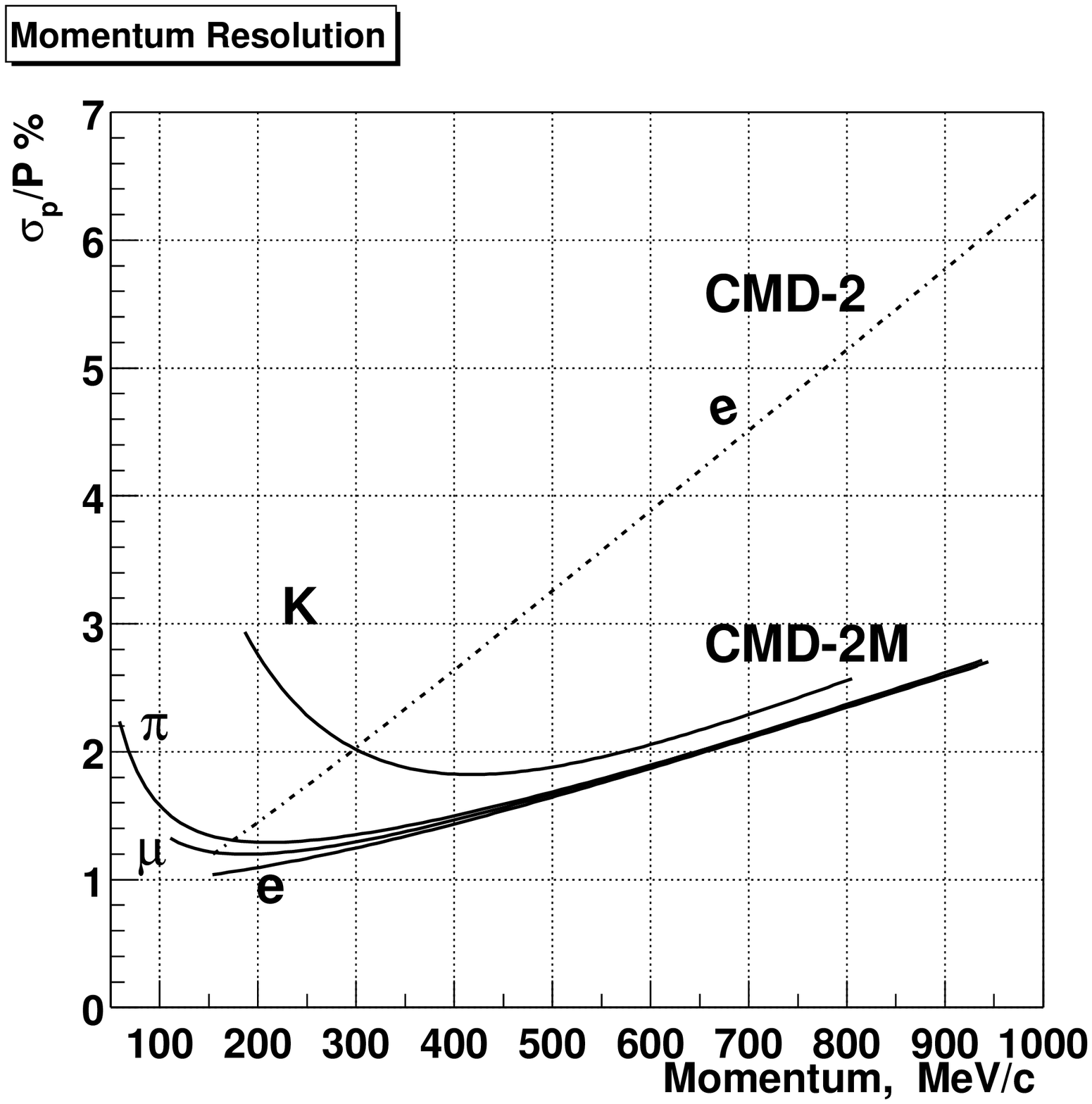}
\end{center}
  \caption{Momentum resolution of the drift chamber.}
\label{fig:dc_res}
\end{figure}

The average spatial resolution in the cell is about 140~$\mu$m
taking into account the cluster effect and diffusion. According
to this spatial resolution and multiple scattering the momentum
resolution is shown on fig.~\ref{fig:dc_res}. 

The Z-coordinate of a track is determined with resolution
of about 2~mm by measurements of charges from both ends of wires. 
Also the measurement of charges is used for dE/dx measurement.
Based on CMD-2 drift chamber experience estimated 
resolution will be 15\%. It allows the separation of kaons 
from pions in a momentum range up to 450~MeV/c. The preamplifiers,
designed for the KEDR detector, are used for
amplification of the signals.
The digitizing electronics are replaced by modern ones.

\section{Z-CHAMBER}

The Z-chamber of the CMD-2 detector is used in the CMD-2M detector.
The Z-chamber is a double layers proportional chamber with
cathode strips~\cite{zc}. The signals from the anode wires are
used for triggering. Signals from cathode strips are used
for measurements of the Z-coordinate of the tracks.

The anode wires with diameter of 28~$\mu$m are made with 
gold-plated \mbox{W-Re}~alloy. The step between wires 
is about 2.8~mm, and the total number
of wires~1408. To minimize the number of readout channels,
the anode wires in both layers are combined in 32 sectors. 
The sectors in the outer layer are rotated relatively 
to the sectors in the inner layer
by half of the angular size of the sectors.  

Use of the fast gas mixture CF$_{4}$:iC$_{4}$H$_{10}$~(80:20)
allows to achieve time resolution of better than 5~ns. Time between
collisions in VEPP-2000 collider is 80~ns. Therefore 
the Z-chamber time resolution is enough for unambiguous
synchronization of events and beams collisions.

There are 256~cathode strips in each layer. The \mbox{Z-co}ordinate
is measured by the centre of gravity of the induced charge method.
High resolution (0.25~mm for a 90~degree track and 1~mm for 
a 45~degree track) allow to perform absolute calibration
of the Z-coordinate measurements in the drift chamber and 
calorimeter.

The upgrade of electronics is only the change. The anode
wires readout is the same with drift chamber. 
The analog electronics used to readout the cathode strips
are of the same design as of those used for 
the liquid Xe calorimeter, while the present ADC cards remain.

\section{ENDCAP CALORIMETER}

The endcap calorimeter covers
forward-backward angles from 16$^o$ to 49$^o$ and from
131$^o$ to 164$^o$. Its total solid angle is 0.3$\times4\pi$.
For the energy of incident photons from 100 to
700 MeV the energy resolution is 8-4\% and the angular resolution 
0.03-0.02 radians.

The endcap calorimeter of the CMD-2 detector consists of 
680~rectangular BGO crystals with
size of \mbox{$25\times25\times150~$mm$^{3}$}, 
arranged in two endcaps~\cite{bgo}.
The thickness of the crystals is 13.4 X$_{0}$ for normal
incidence.
 The transverse dimensions of the crystals,
were chosen as a compromise between the spatial resolution and
the number of electronic channels.
All faces of the crystals are polished and light is collected by
total internal reflection.
The total weight of crystals is about 450~kg.

The light readout is performed by vacuum
phototriodes. Signals from
the phototriodes are amplified by low noise charge sensitive
preamplifiers~\cite{bgo_el}. 
The preamplifiers are placed directly on the
phototriodes inside the detector for best noise
performance. After further amplification and filtration
by shaping amplifiers the signals come to an ADC.

The same BGO crystals are used in the CMD-2M detector but
the photosensitive devices have to be changed. The focusing
solenoids of the VEPP-2000 collider cause nonuniform magnetic
field in the endcap calorimeter volume and limit the available
space from 204~mm to 179~mm. As a result vacuum phototriodes
can not be used. The HAMAMATSU silicon PIN photodiodes with
sensitive area 1$\times$1~cm$^{2}$ look
as a best choice. They are compact and insensitive to the magnetic
field. Additional features of silicon photodiodes are high quantum
efficiency, stability and reliability. 

Silicon photodiodes have an order of magnitude larger capacitance
compare to vacuum phototriodes. So new charge sensitive preamplifiers
have to be designed for better signal-to-noise ratio. The rest
of the electronics does not change. 

Measurements with a prototype show a light yield of 600~e/MeV 
and electronic noise of 500~e. It gives an effective noise 
0.8~MeV which is about the same with CMD-2 data of 0.9~MeV. 
Therefore the expected resolution
of the CMD-2M endcap calorimeter is close to that obtained with
the CMD-2 detector.

\section{BARREL CALORIMETER}

The barrel calorimeter covers polar angles from 38$^o$ to 142$^o$.
Its total solid angle is 0.79$\times4\pi$. It consists of 2~parts:
liquid Xe calorimeter and CsI crystal calorimeter. The thickness
of liquid Xe calorimeter is 5~X$_{0}$. The total volume of liquid Xe
is about 420~l (weight is about 1.2~t). It is a new system in
the CMD-2M detector. The CMD-2 detector barrel calorimeter crystals 
with thickness 8.1~X$_{0}$ are used in the crystal part. 
The number of crystals increases from 892 to 1152  because of
the extension of the internal radius by 15~cm.

\begin{figure}[b]
\begin{center}
\includegraphics[width=0.45\textwidth]{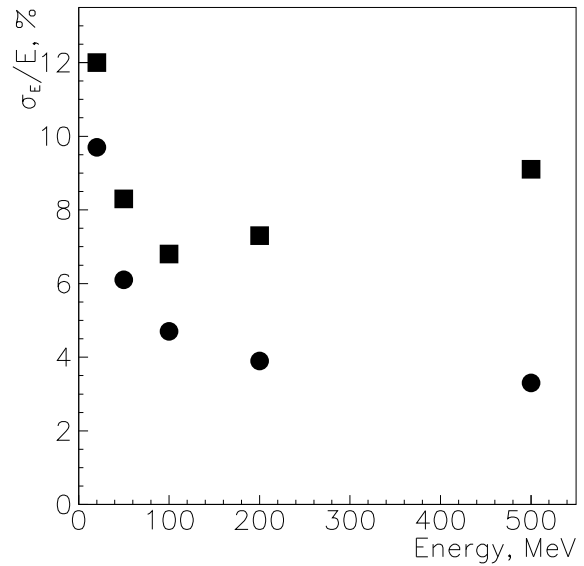}
\end{center}
  \caption{Energy resolution of the barrel calorimeter (MC).
Boxes -- CMD-2, circles -- CMD-2M.}
\label{fig:barrel_res}
\end{figure}

Due to the increase of the calorimeter thickness 
the energy resolution
will be substantially improved~(fig.~\ref{fig:barrel_res}).
The measurement of the coordinates of the conversion points 
will allow to achieve an angular resolution of about 0.005~radians.

\subsection{Liquid Xe (LXe) Calorimeter}

\begin{figure}[tb]
\begin{center}
\includegraphics[width=0.45\textwidth]{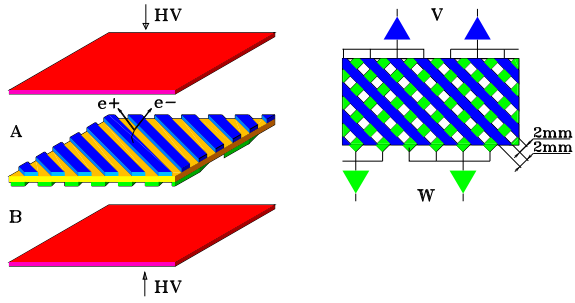}
\end{center}
  \caption{Design of electrodes of the LXe calorimeter.}
\label{fig:lxe_electrodes}
\end{figure}

The LXe calorimeter project is based on successfully carried
R\&D in BINP during the last years~\cite{lxe}. It consists of
a set of ionization chambers with readout from both 
electrodes~(fig.~\ref{fig:lxe_electrodes}). The cathode electrodes
are divided on strips for measurement of conversion
point coordinates. The anode electrodes have rectangular pads
for measurements of an energy deposition. There are 7~cathode
and 8~anode cylindrical electrodes. The longitudinal segmentation
allow to use dE/dx measurements for K$_{L}/\gamma$, 
$\pi^{\pm}/K^{\pm}$ and $\pi$/e separation.
The electrodes are made with copper foil G10 plates with
thickness 0.5~mm~(anode) and 0.8~mm~(cathode). A gap size
is 10.2~mm. Drift time is 4.5~$\mu$s. 

The conversion point coordinates
are measured by the centre of gravity of induced charge method.
Both surfaces of cathode electrodes are divided on 2~mm
width strips spaced by 2~mm. The angle between strips
on different surfaces of one electrode is 90$^{o}$.
This structure is semitransparent: the induced charges on
both surface of electrode is almost the same. It allows
to measure both coordinates of a conversion point in one gap.
This feature is important to measure the angles of
soft photons with short range $e^{+}e^{-}$~conversion pairs.
To minimize the number of readout channels, the strips are grouped
in~4. The total number of cathode channels is~2124. 

The anode pads form towers with projections to the interaction
point. LXe calorimeter is divided onto 8 rings along Z-axis
and onto 33 sectors in R-$\varphi$-plane. So it consists
of 264~towers with angular size of about 11$^{o}\times$11$^{o}$.

The capacitance of the cathodes strips and anode tower are close
to 500~pF. They are readout independently by similar chains.
Analog electronics consist of specially designed 
charge sensitive preamplifiers and shaping amplifiers
and are located inside the detector. The shaping time of
the cathode channels is 4.5~$\mu$s in accordance with
the drift time. The shaping time of the anode channels 
0.4~$\mu$s is chosen as a compromise between the electronic noise
and the geometry factor contributions to the energy resolution.
The electronic noise of the cathode and anode channels is
350~e~+~1.6~e/pF and 900~e~+~2.1~e/pF respectively.
The ADC cards are the same as for the crystal calorimeters.

\subsection{CsI Calorimeter}

The CsI calorimeter of the CMD-2M detector design is based on
experience with  CMD-2 detector barrel calorimeter~\cite{csi}.
It consists of 8~identical octants. Each octant contains 9~linear 
modules~(rows) by 16~crystals. The rows are oriented along
the beam axis. 7~central rows consist of
rectangular crystals with size \mbox{$60\times60\times150~$mm$^{3}$}.
2~side rows are built from special shape crystals to avoid
gaps between octants.

The crystals are covered by diffuse reflector for light collection.
In the CMD-2M detector the light readout from CsI crystals is
performed by HAMAMATSU silicon PIN photodiodes with
sensitive area 1$\times$2~cm$^{2}$ instead of PMTs
as used in the CMD-2 detector. The main reason of changing 
the readout is high sensitivity of the PMTs to the scattered 
magnetic field magnitude which is hard to keep under the control. 

Usage of photodiodes requires more sophisticated electronics
compared to PMT readout. The analog electronics consist of
low noise charge sensitive preamplifiers
designed for endcap calorimeter and upgraded version of designed for 
the KEDR detector shaping amplifiers. The ADC cards remain. 
Measurements with a  prototype show  a light yield of 2500~e/MeV 
and electronic noise of 800~e. It gives an effective noise of 0.3~MeV 
which does not affect the resolution.

\section{RANGE~~SYSTEM}

\begin{figure}[b]
\begin{center}
\includegraphics[width=0.45\textwidth]{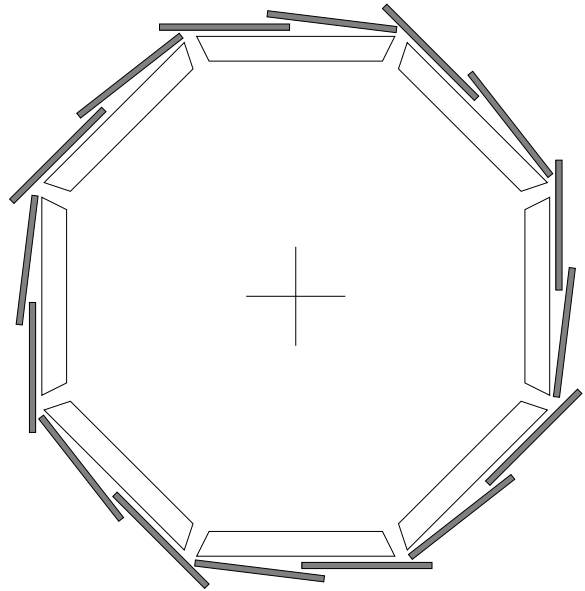}
\end{center}
  \caption{Layout of the range system.}
\label{fig:range}
\end{figure}
The range system of the CMD-2M detector~(fig.~\ref{fig:range})
consists of 16~counters
which are located outside the flux return yoke.  
Beam muons with energy above 550~MeV will reach it at all angles.

The counters are made from plastic scintillator sheets with 
thickness 2~cm. The width and length of the counters is 
40~cm and 150~cm respectively. The scintillator are viewed
from both ends by PMTs. The expected time resolution is better
than 1~ns. It is enough to separate cosmic muons from
beam produced ones.

\section{CONCLUSION}

The project of the upgraded detector CMD-2M for the new
collider VEPP-2000 built at BINP is presented.
 
Compared to the VEPP-2M collider 
the new collider \mbox{VEPP-2000} will have a wider 
centre-of-mass energy range up to 2~GeV and higher
luminosity up to 10$^{32}$~cm$^{-2}$s$^{-1}$.

The general structure of the detector CMD-2 and its
most expensive components, such as BGO and CsI crystals, 
as well as substantial part of electronics
and software will remain as are. At the same time 
all systems of the detector will either be upgraded
or replaced by new ones. After upgrade 
major parameters of the detector,
such as momentum and angular resolutions for 
charged particles and energy and spatial resolutions
for photons, will be improved substantially. 
Improvement of the detector parameters and
upgrade of the accelerator-collider complex
will allow to carry research at a new level
of precision.

\section*{ACKNOWLEDGEMENTS}

Many physicists and engineers from different
institutions are contributed 
to the CMD-2M project and the author is only one of them.
This papers is presented on behalf of the CMD-2M  
collaboration.

The author would like to thank Y.Semertzidis
for reading this paper and making useful remarks.


\newpage

\appendix{\large \bf A CMD-2M COLLABORATION } \label{app}

\begin{center}

R.R.Akhmetshin,
V.M.Aulchenko,
V.Sh.Banzarov,
L.M.Barkov,
S.E.Baru,
N.S.Bashtovoy,
A.E.Bondar,
D.V.Bondarev,
A.V.Bragin,
S.I.Eidelman,
D.A.Epifanov
G.V.Fedotovich,
D.A.Gorbachev,
A.A.Grebenuk,
D.N.Grigoriev,
F.V.Ignatov,
P.M.Ivanov,
S.V.Karpov,
V.F.Kazanin,
B.I.Khazin,
I.A.Koop,
P.P.Krokovny,
E.A.Kuper,
A.S.Kuzmin,
P.A.Lukin,
M.A.Nikulin,
V.S.Okhapkin,
E.A.Perevedencev,
A.S.Popov, 
S.I.Redin,
N.I.Root,
A.A.Ruban,
N.M.Ryskulov,
Yu.M.Shatunov,
B.A.Shwartz,
A.L.Sibidanov,
A.N.Skrinsky,
I.G.Snopkov,
E.P.Solodov,
P.Yu.Stepanov,
A.A.Talyshev,
V.M.Titov,
Yu.Y.Yudin,
S.G.Zverev \\
Budker Institute of Nuclear Physics, 
Novosibirsk, 630090, Russia \\[5mm]
R.M.Carey,  
I.Logashenko,
J.P.Miller,
B.L.Roberts \\
Boston University, Boston, MA 02215, USA \\[5mm]
F.Grancagnolo,
S.Spagnolo \\
Sezione dell'I.N.F.N. di Lecce, 73100, Italy \\[5mm]
J.A.Thompson \\
University of Pittsburgh, Pittsburgh, PA 15260, USA \\[5mm]
S.Dhawan,
V.W.Hughes \\
Yale University, New Haven, CT 06511, USA \\

\end{center}

\end{document}